\documentclass{article}
\usepackage{spconf,amsmath,graphicx}
\usepackage{booktabs}
\usepackage{color,amsmath,url,times, tabularx,bbm,amssymb,multirow}
\usepackage{bm}

\title{Improving Speech Enhancement via Event-based Query}

\name{Yifei Xin$^{1 \ast}$,
      Xiulian Peng$^{2}$,
      Yan Lu$^{2}$\thanks{$^{\ast}$This work was done at Microsoft Research Asia.}
      }
\address{
  $^{1}$Peking University, Beijing, China\\
  $^{2}$Microsoft Research Asia, Beijing, China}
\begin{document}
%
\maketitle
\begin{abstract}
Existing deep learning based speech enhancement (SE) methods either use blind end-to-end training or explicitly incorporate speaker embedding or phonetic information into the SE network to enhance speech quality. In this paper, we perceive speech and noises as different types of sound events and propose an event-based query method for SE. Specifically, representative speech embeddings that can discriminate speech with noises are first pre-trained with the sound event detection (SED) task. The embeddings are then clustered into fixed golden speech queries to assist the SE network to enhance the speech from noisy audio. The golden speech queries can be obtained offline and generalizable to different SE datasets and networks. Therefore, little extra complexity is introduced and no enrollment is needed for each speaker. Experimental results show that the proposed method yields significant gains compared with baselines and the golden queries are well generalized to different datasets. 
\end{abstract}
\section{Introduction}
\label{sec:intro}
Speech enhancement (SE) is a task to improve the intelligibility and perceptual quality of speech signals corrupted with surrounding noises. It has many applications in our life, such as teleconference, mobile phone calls, and automatic speech recognition. Recently, deep learning (DL) based methods have promoted the development of SE methods, as they are more powerful in dealing with non-stationary noise than conventional statistical signal-processing based approaches \cite{wang2018supervised}. Existing DL-based speech enhancement can be classified into time domain \cite{wang2021tstnn,macartney2018improved} and time-frequency (T-F) domain \cite{hu2020dccrn,SN-Net} based methods. Compared with time-domain methods, T-F domain methods are more widely employed as the T-F spectrum matches better with the human perception. They typically predict a mask to filter the noisy T-F spectrogram, thus generating a clean T-F spectrogram. Although these methods have greatly improved the enhanced quality, the recovered speech still suffers from information loss and noise leaking.
\begin{figure}[t]
  \centering
  \includegraphics[width=1.0\linewidth]{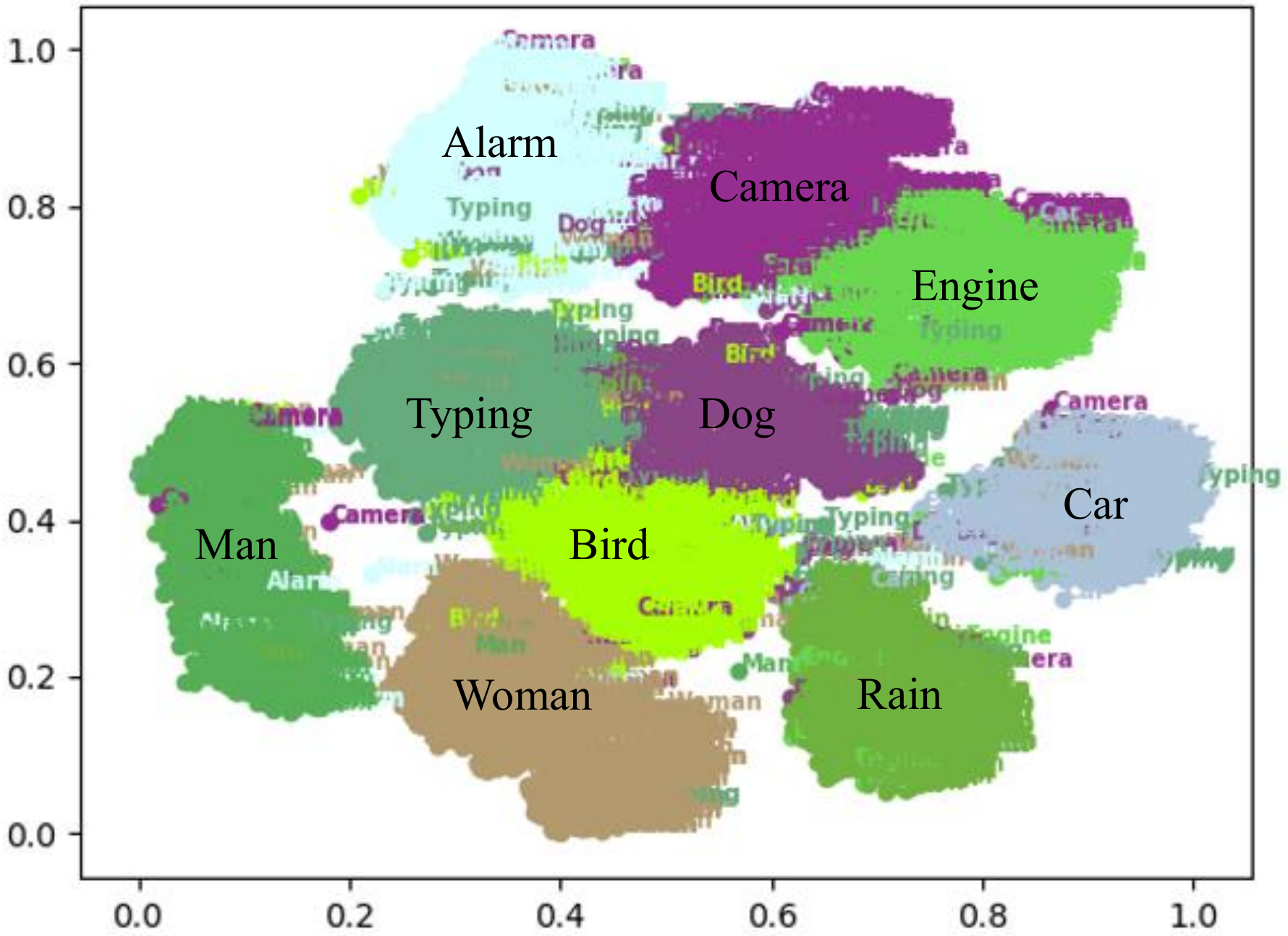}
  \caption{t-SNE visualization of eight noise classes from AudioSet with men and women speech from LibriSpeech based on embeddings from a sound event detection network.}
  \label{fig:t-sne}
  \vspace*{-\baselineskip}
\end{figure}

To further improve the enhancement quality, some methods propose to incorporate semantic information to assist the enhancement process. Chuang et al. \cite{chuang2019speaker} proposed to first train a speaker embedding network to generate the embedding vector for a given speaker, and then utilize the embedding to enhance the target speaker. Liu et al. \cite{liu2021phoneme} proposed to utilize phoneme information to help SE, where distribution modulation parameters learned from phoneme classification vector of the noisy input are used to modulate the SE features. In addition to using information extracted from noisy inputs, Giri et al. \cite{giri2021personalized} proposed to utilize the speaker identity extracted from a clean reference for personalized SE, where the speaker ID is generally embedded into a global vector, and inserted into the reconstruction network. Yue et al. \cite{yue2022reference} proposed a reference-based speech enhancement method by exploring local correlations between noisy and clean reference speech to align and fuse the features of reference and noisy speech together. However, these methods either need an extra online feature extraction process from noisy or clean references with a high cost, or use an enrollment step for each speaker in personalized SE. 

In this paper, we regard speech and noises as different types of sound events and learn offline event-based embeddings to assist SE with negligible extra cost. As shown in Fig. \ref{fig:t-sne}, the embeddings of speech from a sound event detection (SED) network \cite{chen2022hts} can be clustered into ``Man'' and ``Woman'' with good discrimination from different types of noises. We leverage such speech embeddings to learn fixed golden queries to enhance speech from noisy audio in SE networks. Notably, our proposed event-based query method can be flexibly plugged into existing SE networks.

In a nutshell, our contributions are threefold: 
\begin{itemize}
\item To the best of our knowledge, our work is the first to introduce the event-based latent embedding to assist speech enhancement in an offline manner.
\item We analyze the properties of event-based latent embeddings and propose an event-based query method for SE via an attention selection mechanism based on two golden query embeddings with good generalization. The proposed method can be employed on existing SE networks with negligible extra cost.
\item Experimental results on two public datasets, i.e. Voice Bank + DEMAND and Deep Noise Suppression Challenge, demonstrate the superiority of our proposed method. 
\end{itemize}

\section{Proposed method}
\label{sec:proposed method}
\subsection{Overview} 
\begin{figure}[t]
  \centering
  \includegraphics[width=1.0\linewidth]{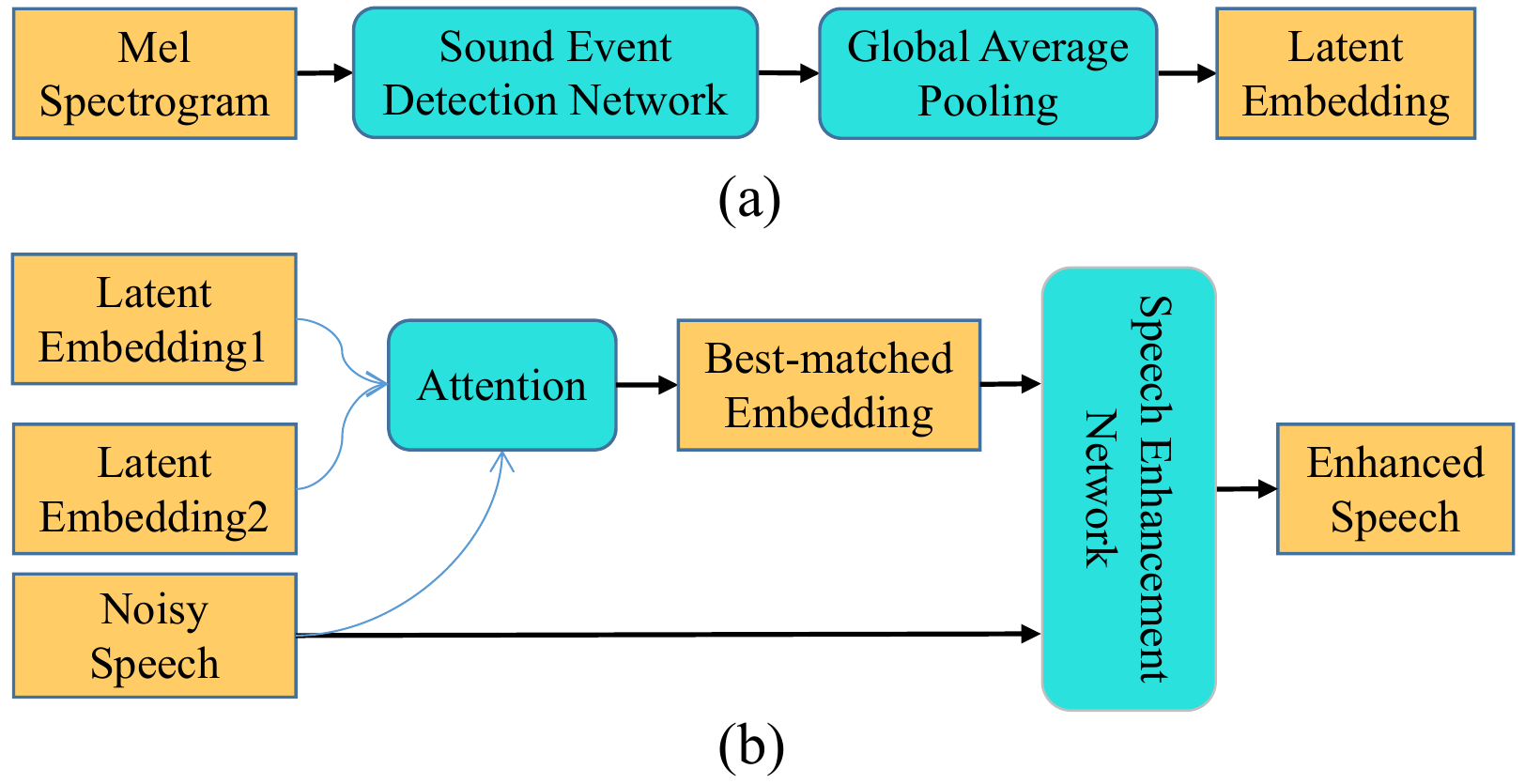}
  \caption{Framework of the proposed method. (a) The sound event detection system. (b) The proposed speech enhancement system.}
  \label{fig:overview}
  \vspace*{-\baselineskip}
\end{figure}

Our framework consists of two parts as shown in Fig. \ref{fig:overview}. First, we learn golden embeddings that can represent speech from a sound event detection (SED) network. Specifically, for each audio clip, we feed it into a SED network to generate the event-based latent embedding. Based on a t-SNE analysis for the latent embeddings, we find that the sound events of people speaking can be clustered according to the gender and speech/noises are clustered into different classes, indicating that the embeddings generated from the SED network can perceive the speaker's gender information and distinguish them from noises. Based on the observations, we try to search two representative golden queries with strong generalization corresponding to gender to assist SE for noisy audio clips. Then, the two golden queries are injected into a speech enhancement network with an attention selection mechanism to choose the best-matched query for each noisy speech. In the following, we will give details for these parts.

\subsection{Sound Event Detection System} 
The Swin Transformer \cite{liu2021swin} structure has been introduced in the audio classification and sound event detection tasks, which achieves competitive performance. In our work, we follow the pipeline in \cite{chen2022hts} to train the Swin Transformer as our backbone of the SED system on AudioSet \cite{gemmeke2017audio} and Librispeech \cite{panayotov2015librispeech} datasets. Specifically, in order to enable the SED network to better perceive the gender information of the speaker, we remove the ``speech'' class of AudioSet, and introduce the Librispeech train-clean-960 dataset. We label each audio of the Librispeech dataset into two categories according to the gender of the speaker, i.e., ``Male speech, man speaking'' and ``Female speech, woman speaking'', which are added as two sound event classes of AudioSet. In this way, the SED model can be trained to pay more attention to the speaker's gender information. 

As show in Fig. 2(a), given an input spectrogram of size $T \times F \times 1$, where $T$ is the number of frames and $F$ is the number of frequency bins, we feed it to the Swin Transformer backbone, which consists of four transformer-encoder blocks. After the four network groups, the shape of the patch tokens is reduced by 32 times to $\frac{T}{32} \times \frac{F}{32} \times D$, where $D$ is the size of latent embedding. Then, we use the global average pooling layer to integrate all time and frequency information into the latent embedding of the shape $(1, D)$. Next, we will visualize the latent embedding to show the SED system's ability to discriminate between different sound events and the perception of the speaker's gender information and then explain how we select the golden queries for speech.

\subsection{Selection of Golden Speech Queries}
As shown in Fig. \ref{fig:t-sne}, we visualize the latent embedding learned from the SED system by t-SNE. We select eight sound event classes from Audioset including ``Car'', ``Bird'', ``Engine'', ``Dog'', ``Rain'', ``Alarm'', ``Typing'', ``Camera''. Besides, we also incorporate 1000 men's and women's clean speech from the Librispeech dataset. From the visualization results, it can be observed that our trained SED system can detect and discriminate different sound events well. More importantly, the SED system can also perceive the gender information of the speaker, and can be aggregated into two clusters. Based on the above results, we attempt to integrate event-based query embeddings into our speech enhancement system. 

To avoid introducing extra complexity during inference, we select two fixed golden queries that can be computed beforehand to represent all speech for the SE system and no online calculation is needed for the embeddings. We choose a diverse multilingual speech dataset to get the two golden query embeddings that can generalize to different datasets. Details of the dataset will be explained in the experimental part. We leverage the mean shift algorithm \cite{comaniciu2002mean} to search the two embeddings with the highest density of ``Man'' and ``Woman'' classes in t-SNE. The densest points of these two classes containing the most information about men and women speakers are chosen as the two golden query embeddings. Based on the selected two embeddings, our event-based query method further incorporates an attention selection mechanism in the SE network to guide it to choose the best-matched (i.e., the same gender) embedding for each audio clip.

\subsection{Speech Enhancement via Event-based Query}
As shown in Fig. \ref{fig:overview} (b), we feed the two selected query embeddings with each noisy audio to the speech enhancement network. For each noisy audio, we adopt the additive attention \cite{bahdanau2014neural} mechanism to calculate the attention score with the two golden embeddings, and then take the one with higher attention score as the best-matched query embedding, i.e., the query of the same gender as the speaker in the noisy audio. After that, the selected best-matched query embedding is fused into every block of the SE network. For each block, the selected embedding is projected to the channel dimension of the corresponding block through a fully connected layer and then added to the feature maps before passing through the next block. In this way, the SE network will learn the relationship between the query embedding and the noisy audio and adjust its weights to adapt to the enhancement of each noisy speech. Both the attention block and SE network are jointly trained from end to end. 

Notably, thanks to the offline calculated golden speech embeddings, our query-based speech enhancement system has only little extra complexity compared to the general speech enhancement system and no enrollment is needed for each speaker. In the experimental part, we will show that the two embeddings are well generalized to different datasets and SE networks.

\section{Experiments and Results}
\label{sec:exp}
\subsection{Datasets}
Two benchmark datasets, i.e., Deep Noise Suppression (DNS) Challenge \cite{dubey2022icassp} and Voice Bank + DEMAND \cite{veaux2013voice}, are used in our experiments. All the utterances are resampled at 16 kHz. 

The DNS challenge at ICASSP 2022 provides a large multilingual speech enhancement dataset, which includes over 2000 speakers from multiple speech corpus and 60000 noise clips with over 150 classes from three noise datasets. We synthesized 890 hours noisy audio with a SNR ranging from -5 to 20 db with reverberations. The SE model performs both noise suppression and dereverberation. We use two test sets for evaluation. One is an internal synthetic dataset (named \textit{multilingual-1458}) that includes 1458 multilingual audio clips, 10 seconds each, without any overlap with the training data. The other is the blind test set of the DNS challenge (named \textit{icassp2022-blind}), which includes 859 real-recorded noisy clips. 

The Voice Bank + DEMAND dataset is a selection of the VoiceBank corpus which includes 11,572 utterances from 28 speakers in the training set and 872 utterances from 2 unseen speakers in the test set. In the training set, the clean utterances are mixed with background noise (8 noise types from DEMAND database and 2 artificial noise types) at SNRs of 0 dB, 5 dB, 10 dB and 15 dB. In the test set, the clean utterances are mixed with 5 unseen noise types from the DEMAND database at SNRs of 2.5 dB, 7.5 dB, 12.5 dB and 17.5 dB. 
\begin{table}
  \caption{Ablation Study on \textit{multilingual-1458} test set.}
  \centering
  \label{tab:freq}
  \begin{tabular}{cccccl}
    \toprule
    Methods & PESQ & CSIG & CBAK & COVL & SSNR\\
    \midrule
    BiDCCRN & 2.72 & 4.15 & 3.36 & 3.48 & 7.87\\
    \midrule
    Ours(ran.l) & 2.75 & 4.14 & 3.32 & 3.46 & 7.91\\
    Ours(ran.) & 2.77 & 4.18 & 3.40 & 3.54 & 7.98\\
    Ours(gold.) & 2.84 & 4.28 & 3.46 & 3.66 & 8.18\\
    Ours(att.) & \textbf{2.91} & \textbf{4.32} & \textbf{3.51} & \textbf{3.72} & \textbf{8.36}\\
    Ours(gen.) & \textbf{2.95} & \textbf{4.36} & \textbf{3.54} & \textbf{3.75} & \textbf{8.45}\\
    Ours(per.) & \textbf{3.02} & \textbf{4.46} & \textbf{3.61} & \textbf{3.80} & \textbf{8.61}\\
  \bottomrule
\end{tabular}
\vspace*{-\baselineskip}
\end{table}
\subsection{Training Details and Evaluation Metrics}
We use two common SE models to evaluate the effectiveness of our method, i.e. Wave-U-Net \cite{macartney2018improved} and BiDCCRN \cite{hu2020dccrn}. We follow the pipeline of Wave-U-Net and BiDCCRN to train the corresponding models. The clean speech of the multilingual-1458 test set are used to get the two golden speech embeddings due to its diversity, no matter the evaluation is done on any other dataset. 

We choose a set of commonly used metrics to evaluate the enhanced speech quality on synthesized test set, i.e., PESQ \cite{rix2001perceptual}, segmental signal-to-noise ratio (SSNR) and mean opinion score (MOS) based metrics: MOS prediction of the signal distortion (CSIG), the intrusiveness of background noise (CBAK) and the overall effect (COVL) \cite{hu2007evaluation}. The non-intrusive evaluation metric DNSMOS P.835 \cite{reddy2022dnsmos} is employed to evaluate the enhanced quality for \textit{icassp2022-blind} without clean targets.
\begin{table}
  \caption{Ablation Study on \textit{icassp2022-blind} with DNSMOS P.835.}
  \centering
  \label{tab:freq}
  \begin{tabular}{cccl}
    \toprule
    Methods & SIG & BAK & OVR\\
    \midrule
    BiDCCRN & 4.22 & 4.41 & 3.90\\
    \midrule
    Ours(gold.) & 4.26 & 4.48 & 3.95\\
    Ours(att.) & \textbf{4.32} & \textbf{4.56} & \textbf{4.02}\\
    Ours(per.) & \textbf{4.40} & \textbf{4.64} & \textbf{4.06}\\
  \bottomrule
\end{tabular}
\vspace*{-\baselineskip}
\end{table}
                           
\subsection{Evaluation on DNS Challenge Dataset}
In this section, we take the bidirectional DCCRN (BiDCCRN) as the SE backbone. The training is performed on the DNS challenge dataset and we evaluate the effectiveness of the proposed method on the \textit{multilingual-1458} and \textit{icassp2022-blind}, respectively. We compare the proposed scheme ``Ours(att.)'' with several baselines. ``BiDCCRN'' denotes the baseline system without event-based query and others denote the event-based query method with various query selection schemes, among which ``Ours(ran.l)'' randomly selects five men and five women's speech from LibriSpeech as candidate queries and for each test audio it randomly selects one from the ten as the additional input of SE network. ``Ours(ran.)'' randomly selects two queries from all the clean targets of the \textit{multilingual-1458} dataset without using the attention mechanism for each test audio. ``Ours(gold.)'' uses the two golden queries but for each test audio it randomly selects one between them as the input of the SE network without using the attention mechanism. ``Ours(gen.)'' and ``Ours(per.)'' are two upperbounds of the proposed scheme that can not be achieved if no extra complexity is allowed, in which the former selects one from two golden queries according to the gender of each test audio instead of using attention mechanism while the latter uses the corresponding clean target of each noisy test audio as query, i.e. we can take it as personalized.  In all these schemes, the same query selection mechanism is used for both training and testing except the last two upperbound schemes which only change the query selection during testing. 

For the \textit{multilingual-1458} dataset, as shown in Table 1, when we randomly select the query embeddings from Librispeech, our query brings negligible improvement or even harms the performance because the Librispeech dataset is not diverse enough and thus it does not generalize well. The random selection of queries is likely to choose the one of the opposite gender, which further harms the enhancement performance. Although random selection is also used in ``Ours(ran.)'' and ``Ours(gold.)'', they show better quality than ``Ours(ran.libri)'', especially ``Ours(gold.)'', which demonstrates the generalization of the two golden queries for speech. The proposed ``Ours(att.)'' outperforms all these random selection schemes and exceeds the ``BiDCCRN'' by a large margin. This shows the effectiveness of both the attention mechanism and the two golden queries. Furthermore, the performance of ``Ours(att.)'' is closed to the ``Ours(gen.)'', showing that the right gender can be found through the attention mechanism. When comparing ``Ours(att.)'' with ``Ours(per.)'', it can be seen that personalized embeddings can further boost the quality with extra cost.

To demonstrate the generalization of our approach, we also present the results on \textit{icassp2022-blind}, the real-recorded noisy dataset in Table 2, where the same two golden queries as the \textit{multilingual-1458} dataset are used here. It is clear that our method ``Ours(att.)'' shows steady performance gains compared with ``BiDCCRN'', showing the good generalization capability of our selected golden queries. Moreover, it is worth noting that our network parameters are only increased by less than 0.5M compared to the corresponding baseline models, which shows the high efficiency of our method.
\begin{table}
  \caption{Comparison on Voice Bank + DEMAND dataset.}
  \centering
  \label{tab:freq}
  \begin{tabular}{ccccl}
    \toprule
    Methods & PESQ & CSIG & CBAK & COVL\\
    \midrule
    Wave-U-Net & 2.40 & 3.52 & 3.24 & 2.96\\
    BiDCCRN & 2.98 & 4.18 & 3.36 & 3.56\\
    \midrule
    Wave-U-Net(att.) & \textbf{2.64} & \textbf{3.68} & \textbf{3.28} & \textbf{3.18}\\
    Wave-U-Net(att.-vb) & \textbf{2.65} & \textbf{3.72} & \textbf{3.28} & \textbf{3.20}\\
    BiDCCRN(att.) & \textbf{3.12} & \textbf{4.30} & \textbf{3.42} & \textbf{3.63}\\
    BiDCCRN(att.-vb) & \textbf{3.16} & \textbf{4.32} & \textbf{3.45} & \textbf{3.66}\\
  \bottomrule
\end{tabular}
\vspace*{-\baselineskip}
\end{table}
\subsection{Evaluation on VoiceBank+DEMAND Dataset}
In this section, we employ the two golden queries obtained from the \textit{multilingual-1458} dataset for evaluation on the VoiceBank+DEMAND dataset. For comparison, we also show the results for golden queries obtained from the training dataset of VoiceBank+DEMAND, denoted as ``att-vb'' in Table 3. Moreover, we evaluate two network structures, BiDCCRN for time-frequency domain methods and Wave-U-Net for time-domain methods to verify the generalization and robustness of our method. As shown in Table 3, our method achieves significant improvements over the corresponding baselines for both SE networks, which demonstrates its generalization capability. Moreover, it can be seen that minor gains are achieved by using golden queries obtained from the training data over that from the \textit{multilingual-1458} dataset, which further shows the generalization capability of the two golden queries selected from the diverse speech dataset \textit{multilingual-1458}.

\section{Conclusions}
\label{sec:conclusion}
In this paper, we propose an event-based query method for SE by exploring the properties of event-based speech query embeddings generated by a SED system. Experiments show that our proposed method can yield significant performance gains compared to the baseline models on multiple datasets using the same golden query embeddings. In the future, we will investigate how to apply our query-based method to more complex scenarios, such as multi-speaker environments.

\bibliographystyle{IEEE.bst}
\bibliography{refs.bib}

\end{document}